\documentstyle[12pt]{article}
\title{Precision Data and Implications on the Parameters of 
TC theory\thanks{ 
This work was supported in part by the National Natural Science 
Foundation of China, and by the funds from 
Henan Science and Technology  Committee.  }}
\author{ Zhenjun Xiao
\thanks{Email: lugr@sun.ihep.ac.cn} \\  
{\small Department of Physics, Henan Normal University,
Xinxiang, 453002 P.R.China.}\\
Shunmin Liu \\
 {\small Department of Physics, Zhumadian Teacher's College,
Zhumadian, 453002 P.R.China.}\\ 
 Xuelei Wang, Lingde Wan and Gongru Lu \\  
{\small Department of Physics, Henan Normal University,
Xinxiang, 453002 P.R.China. }\\ }
\date{\today}

\begin{document}
\maketitle
\begin{abstract}
Assuming that the actual values of $m_t$ 
and the data set  ($\Gamma_b$, $\Gamma_h$, 
$\Gamma_Z$, $R_b$, $R_c$, $R_l$) are within their $1-\sigma$ errors 
as reported by CDF, D0 and by LEP Collaborations, 
the parameter $\Delta^{new}_b$ which 
measures the nonoblique corrections on the $Zb\overline{b}$ vertex 
from new physics can be determined experimentally.
According to the precision data one can obtain updated constraints on the 
parameters $\xi$ and the masses of the charged PGBs.
\end{abstract}
\vspace{1cm}

\newpage

\subsection*{1. Introduction}

As is well known, the Standard Model \cite{Glashow} predictions for the 
electroweak observables are in perfect agreement with the current data 
\cite{Schaile2,Renton}. But frankly speaking, the SM is also a complicated 
theory with many free parameters and other open questions. 
Very recently  the discovery of the top quark has been announced by CDF 
with $M_t=176 \pm 8 \pm 10\;GeV$ \cite{CDF}, which we interprete as 
$M_t=176 \pm 13 \;GeV$,  and by 
the D0 Collaboration with $M_t=199  ^{+19}_{-21} \pm 22\;GeV$\cite{D0}).
This direct measurement of top quark mass is in very good 
agreement with the prediction 
based on the SM electroweak fits of the LEP and other data, 
$M_t=178 \pm 8 ^{+17}_{-18}\;GeV$\cite{Renton}, where the 
central value and the first error refer to $M_H=300\;GeV$. 
This direct measurement of $M_t$, while 
still not very precise, should help in reducing the present uncertainties 
on almost all electroweak observables. And consequently, the knowledge of 
$M_t$ will be very important for one to look for the hints of new physics.

Technicolor(TC) \cite{Farhi} is one of the important candidates for the 
mechanism of electroweak symmetry breaking. 
The comparison of theoretical predictions based on the TC theories 
and the precision electroweak measurements is very specialized and 
rapidly changing, as new data becomes available as well as
new theoretical variables with which theory can be compared 
with  experiment. This subject is of immense importance to TC theory, 
because it has been widely reported that the data disfavor
TC theories, a claim that has been disputed by several authors 
(for a recent review see ref.\cite{King}).

Very recently, Burgess et al., \cite{Burgess} extended the (S,T,U) 
parametrization\cite{Peskin} by introducing three additional parameters 
(V,W,X) 
to describe the lowest non-trivial momentum dependence in oblique 
diagrams. The inclusion of (V, W, X) in the fit 
weakens the bounds on S, T strongly: $S < 2.5$, 
$T < 1.3$ \cite{King,Burgess}.  
 
In this paper  we define a parameter $\Delta_b^{new}$ which 
only measures the non-oblique corrections on 
$Zb\overline{b}$ vertex from new physics, especially that from the 
ETC dynamics and the charged PGBs appeared in QCD-like TC theories.
By the comparison of the theoretical prediction for $\Delta_b^{new}$ in TC 
theory with the experimentally determined $\Delta_{b,exp}^{new}$
one can obtain some constraints on the Clebsch-Gordon coefficient
$\xi$ and put  new lower limits on the masses of charged PGBs.

This paper is organized as follows: In Sec.2 we 
at first present the standard model predictions for $R_b$ and other 
observables and then define the new parameter $\Delta_b^{new}$.
In Sec.3 are collected the relevant calculations and 
the constraints for the parameter $\xi$ and the masses $m_{p1}$ and 
$m_{p2}$ for QCD-like TC theories. We also list and comment on 
several new TC models proposed very recently in the sense of 
avoiding the existed constraints imposed by the precision data. 
The conclusions and the related discussions are in Sec.4.

\subsection*{2. $Zb\overline{b}$ veretx, the SM  predictions and the data}

For LEP processes there are two types of radiative corrections: 
the corrections to the gauge boson self-energies and the corrections 
to the $Zb\overline{b}$ vertex. In  the evaluation 
of self-energy  corrections the error due to  our ignorance of the Higgs mass
is substantial after the direct measurement of $m_t$ at 
Fermilab\cite{CDF,D0}. 
On the other hand, in the corrections to the $Zb\overline{b}$ vertex, 
where the leading contribution due to the large top quark mass is 
produced by the exchange of the W bosons, there is no dependence on the 
unknown Higgs mass. Moreover, the possible new physics contributions 
to the $Zb\overline{b}$ vertex are much more restricted. 
Any non-standard behavior most 
possibly means the existence of new physics!

The Z-pole observables considered in this paper include 
$\Gamma_b$, $\Gamma_h$, $\Gamma_Z$, $R_b$, $R_c$ 
and $R_l$ (in which  $\Gamma_l=(\Gamma_e + \Gamma_\mu + \Gamma_\tau)/3$), 
they are well determined theoretically and experimentally. 
Because the asymmetry $A_{FB}^b$ is almost 
unaffected by the $Zb\overline{b}$ vertex correction 
\cite{Altarelli} we will not include this quantity in our analysis.

Calculations of the one-loop corrections to the $Zb\overline{b}$ vertex
has been performed by several groups 
\cite{Akhundov}. 
The partial decay width $\Gamma (Z\rightarrow f \overline{f})$ has been 
calculated in the $\overline{MS}$ renormalization scheme 
\cite{Degrassi} 
and has been expressed in a compact form 
\cite{Pich},
\begin{eqnarray}
 \Gamma(Z\rightarrow f\overline{f})&=& \frac{N_c^f}{48}\frac{\hat{\alpha}}
{\hat{s}^2_w \hat{c}_w^2}\,m_Z
[\hat{a}_f^2 + \hat{v}_f^2](1+\delta^{(0)}_f)(1+\delta_{QED}^f)\nonumber \\
&&\cdot (1+\delta_{QCD}) (1+\delta_\mu^f)(1+\delta_{tQCD}^f)(1+\delta_b),
\end{eqnarray}
where $N_c^f=3(1)$ for quarks (leptons) is the color factor.  
The partial decay widths in eq.(1) has included the genuine electroweak 
corrections, the QED and QCD corrections, as well as the corrections to 
$Zb\overline{b}$ vertex due to the large top quark mass. 
The definitions and the explicit expressions for all functions and 
factors appeared in eq.(1) can be 
found in refs.\cite{Degrassi,Pich}.  
In ref.\cite{Fleischer}, 
J.Fleischer et al. calculated the two-loop $0(\alpha\alpha_s)$
QCD corrections to the partial decay width $\Gamma_b$, and they found a 
screening of the  leading one-loop top mass effects by $m_t\rightarrow $
$m_t\,[1-\frac{1}{3}(\pi^2-3)\alpha_s/\pi]$. In this paper we will include 
this two-loop QCD corrections. For more details about the calculations 
of $\Gamma_b$ and other relevant quantities in the SM one can see 
the refs.\cite{Akhundov,Degrassi} and a more recent paper\cite{Xiao1}.

In our analysis, the measured values 
\cite{Schaile2,pdg,BES,CDF} 
$m_Z=91.1888$ $\pm 0.0044$ 
$\;GeV$, $G_\mu=1.16639\times 10^{-5}(GeV)^{-2}$ ,
 $\alpha^{-1}=137.0359895$, $\alpha_s(m_Z)=0.125\pm0.005$,
$m_e=0.511\;MeV$, $m_\mu=105.6584\;MeV$ and $m_\tau = 1776.9\;MeV$, 
together with $m_t=176 \pm 13 \;GeV$ and the assumed value 
$M_H=300^{+700}_{-240}\;GeV$ are used as the input parameters.  
In the  numerical calculations
we conservatively take the ``on-shell'' mass of the 
b-quark the value $m_b=4.6\pm 0.3\;GeV$ (in ref.\cite{Pich}, 
the authors used 
$m_b=4.6\pm 0.1\;GeV$), and use the known relation\cite{ckg}between 
the ``on-shell'' and the $\overline{MS}$ schemes to compute the running mass 
$\overline{m}_b(m_Z)$ at the Z scale: 
$\overline{m}_b(m_Z)=3 \pm 0.2\;GeV$ for $m_Z=91.1888\; GeV$. 
We also use the same treatment for the c-quark, 
$\overline{m}_c(m_Z)=1\,GeV$ if we take $m_c=1.6\,GeV$ as 
its ``on-shell'' mass. For other three light quarks we simply assume that 
$\overline{m}_i(m_Z)=0.1\,GeV\;(i=u,d,s)$. 
All these input parameters will
be referred to as the {\em Standard Input Parameters} (SIP).

Among the electroweak observables  
the ratio $R_b=\Gamma_b/\Gamma_h$ is the special one. For this ratio
 most of the vacuum polarization corrections 
depending on the $m_t$ and $m_h$ 
cancel out, while the experimental uncertainties
in the detector response to hadronic events also basically cancel. 
Furthermore, this ratio is also insensitive to extensions of the SM
which would only contribute to vacuum polarizations. 

In Table 1 we list the SM predictions for the Z boson decay widths 
(in MeV) and the ratios $R_b$, $R_c=\Gamma(Z\rightarrow c\overline{c})/
\Gamma_h$ and $R_l=\Gamma_h/\Gamma(Z\rightarrow l\overline{l})$, the 
corresponding measured values at LEP are also listed.  
It is easy to see that the $R_b$ predicted by the SM is smaller than 
that measured. The deviation reaches 2.2-$\sigma$ (or 2.5-$\sigma$ at 
one-loop order) for $m_t=176\;GeV$. 

The precision data can be used to set limits on TC theory 
as well as other kinds of possible new physics. 
Besides the $m_t$ 
dependence the $Zb\overline{b}$ vertex is also sensitive to a 
number of types of new physics. One can parametrize such effects by
\begin{eqnarray}
\Gamma_b=\Gamma_b^{SM}(1+\Delta_b^{new})
\end{eqnarray}
where the term $\Delta_b^{new}$ represents the pure 
non-oblique corrections 
to the $Zb\overline{b}$ veretx from new physics.
The partial decay width $\Gamma_b^{SM}$ can be determined 
theoretically by eq.(1),  
and consequently other five observables studied in this paper 
can be written as the form of 
\begin{eqnarray}
\Gamma_h&=&\Gamma_h^{SM}+\Gamma_b^{SM}\cdot \Delta_b^{new}, \ \ \ \ 
\Gamma_Z=\Gamma_Z^{SM}+\Gamma_b^{SM}\cdot \Delta_b^{new},\nonumber\\ 
R_b&=&R_b^{SM}+R_b^{SM}(1-R_b^{SM})\cdot \Delta_b^{new},\ \ \ \  
R_c=R_c^{SM}-R_b^{SM}R_c^{SM}\cdot \Delta_b^{new},\nonumber\\ 
R_l&=&R_l^{SM}+\frac{\Gamma_b^{SM}}{\Gamma_l^{SM}}\cdot \Delta_b^{new}.
\end{eqnarray}
Obviously, the oblique corrections and the 
heavy top quark vertex effect  have been absorbed into the 
evaluations for the observables $X_i^{SM}$ in the SM.  This definition of 
$\Delta_b^{new}$ in eq.(2) is different from that of $\epsilon_b$ 
\cite{Altarelli}(as well as the parameter $\Delta_b$ in 
refs.\cite{Blondel,Cornet}). 
In the SM the parameters $\epsilon_b$\cite{Altarelli} 
and $\Delta_b$\cite{Cornet} are closely 
related to the quantity $-Re\{\delta_{b-vertex} \}$ defined in 
ref.\cite{Pich} and are dominated by quadratic terms in $m_t$ of order 
$G_F m_t^2$. While the parameter $\Delta_b^{new}$ only measures the 
new physics effects on the $Zb\overline{b}$ vertex, and 
$\Delta_b^{new} \equiv 0$ 
in the SM. We think that this definition of $\Delta_b^{new}$ is more 
convenient than other similar definitions to measure the new physics 
effects on the $Zb\overline{b}$ vertex, 
since new physics can  be disentangled if not 
masked by large $m_t$ effects.  

In order to extract the vertex factor $\Delta_b^{new}$ from the data set 
$(\Gamma_b, \Gamma_h, \Gamma_Z, R_b, R_c, R_l)$ as listed in Table 1  
more quantitatively, 
we construct the likelihood function of  $\Delta_b^{new}$ 
as the form of 
\begin{eqnarray}
{\cal L}(x_{exp}, \Delta_b^{new})= N\,Exp[-\sum_x \frac{1}{2}
(\frac{x_{exp}-x( \Delta_b^{new})}{\sigma_x})^2] 
\end{eqnarray} 
where the $\sigma_x$ is the experimental error of the observable 
$ x_{exp}$, and N is the normalization factor. 
With the SIP, the point which maximizes 
${\cal L}(x_{exp}, \Delta_b^{new})$ is found to be 
$\Delta_b^{new} = 0.001$ for $m_t=176\; GeV$. And we also have 
\begin{eqnarray}
\Delta_b^{new} = 0.001 \pm 0.005
\end{eqnarray}
at $1-\sigma$ level for $m_t=176\;GeV$ and $M_H=300\;GeV$, 
while the remainder uncertainties of 
$\Delta_b^{new}$ are $\pm 0.002$ and $ ^{+0.004}_{-0.002}$ corresponding to 
$\delta m_t=13\,GeV$ and $M_H=300^{+700}_{-240}$ respectively.
It is easy to see that $\Delta_b^{new}$ is now consistent with zero at 
$1-\sigma$ level. 
By its own definition the parameter $\Delta_b^{new}$ has no dependence 
on $M_H$, the present weak dependence is coming from the standard model 
calculations for the six observables. In the following 
analysis we always use $(m_t=176\,GeV$, $M_H=300\,GeV$) as 
reference point and don't discuss the variation of $M_H$.

If we interpret the quantity
\begin{eqnarray}
P(\Delta_b^{new} > A) = \int_A^{+\infty}
d \Delta_b^{new} {\cal L}(x_{exp}, \Delta_b^{new}) 
\end{eqnarray}
and 
\begin{eqnarray}
P(\Delta_b^{new} < B) = \int^B_{-\infty}
d \Delta_b^{new} {\cal L}(x_{exp}, \Delta_b^{new})
\end{eqnarray}
as the probability that $\Delta_b^{new} > A$ ($\Delta_b^{new} < B$ ), 
then one can  obtain the $95\%$ one-sided upper (lower) confidence 
limits on $\Delta_b^{new}$:
\begin{eqnarray}
\Delta_{b,exp}^{New} > -0.010,\ \ and \ \ 
\Delta_{b,exp}^{New} < 0.012 
\end{eqnarray}
for $m_t=176\pm 13\;GeV$. 

For any kinds of new physics which may contribute to the $Zb\overline{b}$ 
vertex, they should satisfy this constraint from $Zb\overline{b}$
vertex as well as those from the (S, T, U, V, W, X) oblique parameters 
simultaneously. 

\subsection*{3. Updated constraints on $\xi$ and masses of charged PGBs}

In the TC models \cite{Farhi,King,Weinberg}, 
the larger top quark mass is presumably the result 
of ETC \cite{Susskind} dynamics at relatively low energy scales. 
There are two sources of corrections to this $Zb\overline{b}$ vertex
in TC models, namely from ETC gauge boson exchange \cite{Simmons,Chivukula} 
and from charged PGB exchange \cite{Xiao2,Xiao3}

For the One-Doublet Technicolor Model(ODTM)\cite{Dimopoulos}, 
no Pseudo-Goldstone bosons can be survived when the 
chiral symmetry was broken by the condensate $<T\overline{T}> \neq 0$, 
but the ETC gauge boson exchange can produce typically large and 
negative contributions to the  $Zb\overline{b}$ vertex, 
as described in ref.\cite{Simmons},  
\begin{eqnarray}
\Delta_1^{ETC} \approx -6.5\%\times \xi^2\cdot [\frac{m_t}{176GeV}]
\end{eqnarray} 
where the constant $\xi$ is an ETC-gauge-group-dependent Clebsch-Gordon 
coefficient and expected to be of order 1 \cite{Simmons}. 
Theoretically, the exact value of $\xi$ will 
be determined by the choice of ETC gauge group and by the assignments of 
the technifermions. As shown in eq.(9), the non-oblique correction
on the $Zb\overline{b}$ vertex from the ETC dynamics is quadratic in $\xi$. 
The variation of $\xi$ will strongly affect the size of $\Delta_1^{ETC}$.
Naturally the experimental limits on the vertex factor $\Delta_b^{new}$ cab be 
interpreted as the bounds on $\xi$.
For $m_t=189\;GeV$ one can have,
\begin{eqnarray}
\xi < 0.4,  \ \ at\ \ 95\%\,C.L.
\end{eqnarray} 
For lighter top quark this bound will be loosened slightly. 

In the most frequently studied Farhi-Susskind One Generation Technicolor 
Model (OGTM) \cite{Dimopoulos}, the global flavor symmetry $SU(8)_L\times 
SU(8)_R$ will break down to the $SU(8)_V$ by technifermion condensate
$<\overline{T}T>\neq 0$. And consequently 63 massless (Pseudo)-Goldstone bosons will be produced from this breaking. 
Besides the nonoblique corrections $\Delta_2^{ETC}$ from the 
ETC gauge boson exchange, 
the charged PGBs in the OGTM also contribute a negative correction 
to the $Zb\overline{b}$ vertex as estimated in ref.\cite{Xiao2,Xiao3}.
In short, 
\begin{eqnarray}
\Delta_b^{new}(OGTM) = \Delta_2^{ETC} + \Delta_b^{P^\pm} 
+ \Delta_b^{P_8^\pm}.
\end{eqnarray} 
where the terms  $\Delta_b^{P^\pm}$ and $\Delta_b^{P_8^\pm}$ represent 
the contributions from the color singlet charged PGBs $P^{\pm}$
and the color octets $P_8^{\pm}$. 
Specifically, all three terms in the right-hand side of this equation 
are  negative. 

For simplicity, 
we  assume that the ETC part of the OGTM studied here are the same or very 
similar with the ODTM studied in ref.\cite{Simmons} except for the difference 
in the value of $F_\pi$ (in the OGTM, $F_\pi= 123\;GeV$), and then we can
write  
\begin{eqnarray}
\Delta_2^{ETC}  \approx -12.9\%\times \xi^2\cdot [\frac{m_t}{176GeV}]
\end{eqnarray} 
Typically, $\Delta_2^{ETC}\approx -6.5\%$ for $m_t=176\;GeV$ and 
$\xi=1/\sqrt{2}$, which is consistent  with the result 
as shown in the Fig.3 of ref.\cite{Chivukula} for the $SU(4)_{ETC} 
\rightarrow$ $SU(3)_{TC}$ model with a full family of technifermions. 

In ref.\cite{Xiao2,Xiao3}, 
we have calculated the non-oblique corrections on the 
$Zb\overline{b}$ vertex from the color singlet PGBs $P^\pm$ and 
the color octet
PGBs $P_8^{\pm}$ respectively. 
The size of the  vertex factor $\Delta_{b}^{P^\pm}$ ( 
$\Delta_{b}^{P_8^\pm}$) depends on $m_t$ and $m_{p1}$ ($m_{p2}$). 
Using the SIP, one can estimate the ranges of the  
term $\Delta_b^{P^\pm}$ and  
$\Delta_b^{P_8^\pm}$:
\begin{eqnarray}
\Delta_b^{P^\pm} &=& (-0.013 \sim -0.002),\; for \ \ 
m_{p1}=50 - 400 \;GeV, \\ 
\Delta_b^{P_8^\pm} &=& (-0.050 \sim -0.003),\; for \ \ 
m_{p2}=200 - 650 \;GeV,
\end{eqnarray} 
where $m_{p1}$ is the mass of 
$P^\pm$, and $m_{p2}$ is the mass of $P_8^\pm$. 
The contributions from the charged PGBs are always negative and will push 
the OGTM prediction for $\Delta_b^{new}$ away from the 
measured $\Delta_{b,exp}^{New}$ to a high degree. 
These negative corrections are clearly disfavored by the current 
data. But fortunately, the charged PGBs show a clear decoupling 
behavior as listed in eqs.(13, 14).  

In the OGTM, the size of vertex factor $\Delta_b$ generally depend on three
" free" parameters, the Clebsch-Gordon coefficient $\xi$, the masses $m_{p1}$
and $m_{p2}$ if we use $m_t=176\pm 13$ GeV as input. In order to study the 
nonoblique corrections on the  $Zb\overline{b}$ vertex more quantitatively, 
we consider the following two ultimate cases:
(a). Under  the limit $\xi \rightarrow 0$, to extract the possible bounds 
on the masses of $m_{p1}$ and $m_{p2}$; 
(b). Under the limits $\Delta_b^{P^{\pm}}\rightarrow 0$ and 
$\Delta_b^{P_8^{\pm}}\rightarrow 0$( e.g. the charged PGBs are heavy enough
and decoupled from the low energy physics), to extract the bounds on the 
parameter $\xi$. 
 
At first if we set $\xi \rightarrow 0$ the current data 
will permit us to exclude large part of the ranges 
of $m_{p1}$ and  $m_{p2}$ in the $m_{p1}-m_{p2}$ plan, the updated bounds on 
the masses of charged PGBs are the following:
\begin{eqnarray}
m_{p1} > 200\;GeV\ \ at\ \  95\%\;C.L., \ \ for\ \ ``free''\ \ m_{p2}
\end{eqnarray}
and 
\begin{eqnarray}
m_{p2} > 600\;GeV\ \ at \ \ 95\%\;C.L., \ \ for \ \ m_{p1} \leq \;400\; GeV.
\end{eqnarray} 
while the uncertainties of $m_t$, $\delta m_t=13\;GeV$, almost don't 
affect the constraints.  These limits are much 
stronger than that has been given before in ref.\cite{Xiao3}.  Of cause, the 
inclusion of the negative corrections from ETC dynamics in the OGTM will
strengthen the bounds on $m_{p1}$ and $m_{p2}$.

Secondly, if we set the limits $\Delta_b^{P^{\pm}}\rightarrow 0$ and 
$\Delta_b^{P_8^{\pm}}\rightarrow 0$,  the current data means 
a stringent bound on the size of $\xi$ in the OGTM:  
$\xi < 0.28$ at $95\%\;C.L.$ for $m_t = 189\;GeV$.
If the charged PGBs are heavy and decoupled and, at the same time,  
the coefficient $\xi$ in QCD-like TC models  can be reduced 
to $0.28$ instead of the popular size $1/\sqrt{2}$ as used 
in ref.\cite{Chivukula}, the magnitude of both the 
$\Delta_{b}^{new}(ODTM)$ and $\Delta_b^{new}(OGTM)$
will be consistent with the present constraints on $\Delta_b^{new}$.

\subsection*{4. Conclusions}

As mentioned at the beginning, TC theory can provide a natural, dynamical 
explanation for electroweak symmetry breaking. But, 
as is well known, this theory (including the ETC) also encountered many 
problems as discussed in detail in refs.\cite{King}. 
At present, the situation becomes better than 3 years ago\cite{Lane}.
The experimentally determined parameters $S_{exp}$ and $\Delta_{b,exp}^{new}$ 
are all close to zero with small errors, and therefore the former strong 
constraints are now weakened.

In ref.\cite{Chivukula}, the authors have shown that a slowly 
running technicolor 
coupling will affect the size of non-oblique corrections to the 
$Zb\overline{b}$ vertex from ETC dynamics. Numerically, the ``Walking TC''
\cite{Holdom} reduces the magnitude of the corrections at about $20\%$ 
level. Although this 
decrease is helpful to reduce the discrepancy between the TC models and the 
current precision  data, however, this improvement is not large enough 
to resolve this problem. More recently, N.Evans\cite{Evans} points out that 
the constraints from $Zb\overline{b}$ veretx may be avoided if the ETC scale $M_{ETC}$ 
can be boosted by strong ETC effects. 

For standard ETC dynamics\cite{Susskind,King} 
the ETC gauge bosons are the $SU(2)_w$ singlets, 
and the exchanges of such kinds of ETC 
gauge bosons will produce large negative corrections to the $Zb\overline{b}$
vertex as described in refs.\cite{Simmons,Chivukula}. In ``Non-commuting''
theories ( i.e., in which the ETC gauge boson which generates the top quark 
mass does carry weak SU(2) charge), as noted in 
refs.\cite{Simmons,Chivukula2}, 
the contributions on the $Zb\overline{b}$ vertex come from the 
physics of top-quark mass generation and from weak gauge boson mixing
(the signs of the two effects are opposite)\cite{Chivukula2}, 
and therefore both the size and the sign of the corrections 
are model dependent and the overall effect may be small and may even 
increase the $Zb\overline{b}$ branching ratio.  It is important 
to explore this class of models further, since the experiments favor 
a larger $R_b$\cite{Xiao1}.

Besides the new TC models just mentioned above  several  TC models with 
novel ideas have also been constructed since 1993, such as the ``Low-scale
technicolor''\cite{King2}, the ``Technicolor model with a scaler''
\cite{Georgi2}, the `` Topcolor assisted technicolor''
\cite{Hill}, ``Chiral technicolor''\cite{Terning2} and other models.
The main motivation for constructing these new models is evident: 
Generating the larger top quark mass and at the same time being  
consistent with the precision data.

In summary we defined a parameter $\Delta_b^{new}$ which 
measures the non-oblique corrections on the 
$Zb\overline{b}$ vertex from the new physics, such as the  
ETC dynamics and the charged PGBs appeared in QCD-like TC theories.
By its own definition the parameter $\Delta_b^{new}$ is different from
the $\epsilon_b$ and the $\Delta_b$ as defined in refs.
\cite{Altarelli,Blondel}, and this parameter can be 
determined experimentally from the data set ($\Gamma_b$, $\Gamma_h$, 
$\Gamma_Z$, $R_b$, $R_c$, $R_l$). 
By the comparison of the theoretical 
prediction for $\Delta_b^{new}$ in QCD-like TC 
theories with the experimentally determined $\Delta_{b,exp}^{new}$
one can obtain some constraints on the Clebsch-Gordon coefficient
$\xi$ and put  more stringent lower limits on the masses of charged PGBs.
From the numerical calculations and the phenomenological analysis 
we found that: 

(a). The charged  Pseudo-Goldstone bosons must be 
heavier than that estimated before in Ref.\cite{Xiao2}. At present
for $m_t=176\pm 13\;GeV$, we have 
$m_{p1} > 200\;GeV$  at $95\% C.L$ for ``free'' $m_{p2}$, 
and $m_{p2} > 600\;GeV$  at $95\% C.L$ 
for  $m_{p1}\leq 400\;GeV$;

(b). If the charged PGBs are indeed very heavy and decoupled and, 
at the same time, the coefficient $\xi$ in the new QCD-like TC models can 
be smaller than 0.28, such kinds of QCD-like  TC models still be allowed.

(c). There is definite discrepancy about the value of $R_b$ between the 
SM and the experiment. But at present 
it is hard to explain this deviation as a signal of new physics. 
From the data set of 
$(\Gamma_b, \Gamma_h, \Gamma_Z, R_b, R_c, R_l)$, one can determine the 
size of the nonoblique corrections on the $Zb\overline{b}$ vertex 
from the new physics experimentally: $\Delta_{b,exp}^{new}=0.001\pm 
0.005\pm 0.002(m_t)$, which is close to zero with small errors.

\newpage
Table 1. The SM predictions for the observables  
$(\Gamma_b,\; \Gamma_h,\; \Gamma_Z,\;R_b,\;R_c,\;R_l)$, 
compared with the measured Z parameters at LEP.
\begin{center}
\vspace{0.2cm}
\begin{tabular}{c|l|l} \hline\hline
 &  SM Predictions& LEP Values \\ \hline
$\Gamma_b$& $377.7 \pm 0.2(m_t) ^{+0.2}_{-0.9}(m_h) 
\pm 0.5(\alpha_s)  \pm 0.4(\hat{\alpha}) \pm 0.3(\overline{m}_b)$ 
& $382.7\pm 3.1$, \cite{Altarelli} \\ \hline 
$\Gamma_h$& $1749.3\; \pm 3.2(m_t)\; ^{+1.4}_{-4.5}(m_h)\; 
\pm 2.9(\alpha_s) \; \pm 1.7(\hat{\alpha})\; \pm 0.3(\overline{m}_b)$ 
& $1745.9\pm 4.0$, \cite{Schaile2} \\ \hline 
$\Gamma_Z$& $2503.9\; \pm 4.3(m_t)\; ^{+1.2}_{-5.9}(m_h)\; 
\pm 2.9(\alpha_s) \; \pm 2.4(\hat{\alpha})\; \pm 0.3(\overline{m}_b)$ 
& $2497.4\pm 3.8$, \cite{Schaile2} \\ \hline 
$R_b$&  $ 0.2159 \pm 0.0005(m_t) \pm 0.00003(m_h) 
\pm 0.00004(\alpha_s)$ $ \pm 0.0001(\overline{m}_b)$, 
& $0.2202\pm 0.0020$, \cite{Schaile2} \\ \hline 
$R_c$& $0.1721\; \pm 0.0002(m_t)\; \pm 0.00004(m_h)\; 
\pm 0.0001(\alpha_s)$$ \pm 0.00003(\overline{m}_b)$, 
& $0.1583\pm 0.0098$,\cite{Schaile2}\\ \hline   
$R_l$&  $20.820 \pm 0.002(m_t) \pm 0.015(m_h) 
\pm 0.034(\alpha_s) \pm 0.003(\overline{m}_b) $ 
& $20.795\pm 0.040$, \cite{Schaile2}\\ \hline \hline  
\end{tabular}
\end{center}


\newpage

\begin{thebibliography}{99}

\bibitem{Glashow}
 S.L.Glashow, {\it Nucl.Phys.} 22(1961)579;\\
 A.Salam, {\em in Elementary Particle Theory,} ed.  
N.Svartholm(Stockholm, 1968);\\
 S.Weinberg, {\it Phys.Rev.Lett.} 19(1967)1246.

\bibitem{Schaile2}
D.Schaile, {\it Precision Tests of the Electroweak Interaction}, Talk 
given at the 27th International Conf. on High Energy Physics, Glasgow, 
20-27th July 1994, CERN-PPE/94-162.  

\bibitem{Renton}
Peter B. Renton, {\it Review of Experimental Results on Precision Tests of 
Electroweak Theories}, Invited talk given at the 17th International 
Symposium on Lepton-Photon Interactions, August 10-15, 1995, Beijing, China,
CERN-PPE/96-63.

\bibitem{CDF}
 CDF Collaboration,  F.Abe et al., {\it Phys.Rev.Lett.} 73(1994)225; \\ 
{\it Phys.Rev.Lett.} 74(1995)2626.  

\bibitem{D0}
D0 Collaboration, S.Abachi et al., {\it Phys.Rev.Lett.} 74(1995)2632.  

\bibitem{Farhi}
E.Farhi and L.Susskind, {\it Phys.Rep.} {\bf 74}(1981)277;\\  
R.K.Kaul,{\it Rev.Mod.Phys.} {\bf 55}(1983)449, and reference there in.

\bibitem{King}
S.F.King, {\it Rep.Prog.Phys.} {\bf 58}(1995)263. 

\bibitem{Burgess}
N.Evans, {\it Phys. Rev.} D{\bf 49}(1994)4785; 
C.P.Burgess, S.Goldfry, M.Konig, D.London and I. Maksymyk,  
{\it Phys.Lett.} {\bf 326B}(1994)276; 
I.Maksymyk, C.P.Burgess and  D.London,     
{\it Phys.Rev.} D{\bf 50}(1994)529.

\bibitem{Peskin}
M.E.Peskin and T. Takeuchi, {\it Phys.Rev.Lett.}  {\bf 65}(1990)964; \\
{\it Phys.Rev.} {\bf D43}(1992)381

\bibitem{Altarelli}
G.Altarelli, R.Barbieri and F.Caravaglios,  
{\it Nucl.Phys.} {\bf B405}(1993)3; CERN-TH.7536/94.

\bibitem{Akhundov}
A.A.Akhundov, D.Yu.Bardin and T.Riemann, {\it Nucl.Phys.} {\bf B276}(1986)1;\\
 J.Bernab$\acute{e}$u, A.Pich and A. Santamaria,
{\it Phys.Lett.} {\bf B200}(1988)569;\\
W.Beenakker and W.Hollik, {\it Z.Phys.} {\bf C40}(1988)141; \\
B.W.Lynn and R.G.Stuart, {\it Phys.Lett.} {\bf B252}(1990)676.

\bibitem{Degrassi}
G.Degrassi and A.Sirlin,  {\it Nucl.Phys.} {\bf B351}(1991)49;

\bibitem{Pich}
 J.Bernab$\acute{e}$u, A.Pich and A. Santamaria,
{\it  Nucl.Phys.} {\bf B363}(1991)326.

\bibitem{Fleischer}
J.Fleischer, O.V.Tarasov  and F.Jegerlehner,  {\it Phys.Lett.} 
{\bf 293B}(1992)437.

\bibitem{Xiao1}
Zhenjun Xiao, Lingde Wan, Gongru Lu and Xuelei Wang, 
 {\it J.Phys.G:} {\it Nucl. Part. Phys.} {\bf 21}(1995)167

\bibitem{pdg}
Particles Data Group, L. Montanet et al., {\it Phys.Rev.} D{\bf 45}(1994)1173.
\bibitem{BES}
BES Collab., J.Z.Bai et al., {\it Phys.Rev.Lett.} {\bf 69}(1992)3021.

\bibitem{ckg}
K.G.Chetyrkin and J.H.Kuhn, {\it Phys.Lett.} {\bf 248B}(1990)359.

\bibitem{Blondel}
 A.Blondel and C.Verzegnassi, {\it Phys.Lett.} {\bf 311B}(1993)346; \\
 A.Blondel, A.Djouadi and C.Verzegnassi, {\it Phys.Lett.} {\bf 293B}
(1992)253. 

\bibitem{Cornet}
F.Cornet, W.Hollik and M.M $\ddot{o}$sle,  
{\it Nucl.Phys.} {\bf B428}(1994)61; 

\bibitem{Weinberg}
S.Weinberg, {\it Phys.Rev.} D{\bf 13} 974(1976); D{\bf 19}(1976)1277;\\
L.Susskind, {\it Phys.Rev.} D{\bf 20}(1979)2619.

\bibitem{Susskind}
 S.Dimopoulos and L.Susskind, {\it Nucl.Phys.} {\bf B155}(1977)237; \\
 E.Eichten and K.Lane, {\it Phys.Lett.} {\bf 90B}(1980)125. 

\bibitem{Simmons}
 R.S.Chivukula, S.B.Selipsky and E.H.Simmons,{\it Phys.Rev.Lett.}
 {\bf 69}(1992)575. 

\bibitem{Chivukula}
 R.S.Chivukula, E.Gates, E.H.Simmons and J.Terning, 
{\it Phys.Lett.} {\bf 311B}(1993)157.

\bibitem{Xiao2}
Zhenjun Xiao, Lingde Wan, Jinmin Yang and Gongru Lu,  
{\it Phys.Rev.}  D{\bf 49} (1994)5949

\bibitem{Xiao3}
Zhenjun Xiao, Lingde Wan, Gongru Lu, Jinmin Yang, Xuelei Wang, Lipuo Guo 
and Chongxing Yue, {\it J.Phys.G:} {\it Nucl.Part.Phys.} {\bf 20}(1994)901

\bibitem{Dimopoulos}
 S.Dimopoulos, {\it  Nucl.Phys.} {\bf B168}(1980)69;\\
 E.Farhi and L.Susskind, {\it Phys.Rev.} D{\bf 20}(1979)3404;\\  
 S.Dimopoulos et al., {\it Nucl.Phys.} {\bf B176}(1980)449.

\bibitem{Lane}
K.Lane, {\it Technicolor},  hep-ph/9501249.


\bibitem{Holdom}
B.Holdom, {\it Phys.Lett.} {\bf 105B}(1985)301; \\
T.Appelquist, D.Karabali and L.C.R. Wijewardhana,   
 {\it Phys.Rev.} D{\bf 35} (1987)389;  \\
T.Appelquist  and L.C.R.Wijewardhana, {\it Phys.Rev.} D{\bf 35}(1987)774; \\
T.Appelquist  and G.Triantaphyllou, {\it Phys.Lett.} {\bf 278B}(1992)345.

\bibitem{Evans}
N.Evans, {\it Phys. Lett.} {\bf 331B}(1994)378.

\bibitem{Chivukula2}
R.S.Chivukula, E.H.Simmons  and J.Terning, {\it Phys. Lett.} 
{\bf 331B}(1994) 383;\\
D.B.Kaplan, {\it Nucl.Phys.} {\bf B365}(1991)259.

\bibitem{King2}
S.F.King, {\it Phys. Lett.} {\bf 314B}(1993)364.

\bibitem{Georgi2}
C.D.Carone  and H.Georgi, {\it Phys.Rev.} D{\bf 49}(1994)1427; 
C.D.Carone, E.H.Simmons and Yumian Yu, {\it Phys.Lett.} {\bf 344B}
(1995)287, and reference therein.

\bibitem{Hill}
C.T.Hill,  {\it Phys. Lett.} {\bf 345B}(1995)483.

\bibitem{Terning2}
J.Terning, {\it Phys. Lett.} {\bf 344B}(1995)279.

\end{thebibliography}
\end{document}